\magnification=\magstep1
\tolerance=500
\rightline{TAUP 2550-99}
\rightline{22 June, 2000}
\vskip 2 true cm
\centerline{\bf Classical Radiation Reaction Off-Shell}
\centerline{\bf Corrections to the Covariant Lorentz Force}
\bigskip
\centerline{O. Oron and L.P. Horwitz\footnote{*}{Also at Department of
Physics, Bar Ilan University,  Ramat Gan 529000, Israel}}
\smallskip
\centerline{School of Physics and Astronomy}
\centerline{Raymond and Beverly Sackler Faculty of Exact Sciences}
\centerline{Tel Aviv University, Ramat Aviv 69978, Israel}
\bigskip

\noindent
{\it Abstract:\/}
\par It has been shown by Gupta and Padmanabhan that the radiation
reaction force of the
Abraham-Lorentz-Dirac equation can be obtained by a coordinate
transformation from the inertial frame of an accelerating charged
particle to that of the laboratory. We show that the problem may be
formulated in a flat space of five dimensions, with five corresponding
gauge fields in the framework  of the
classical version of a fully gauge covariant form of the 
Stueckelberg-Feynman-Schwinger covariant
mechanics (the  zero mode fields of the $0,1,2,3$ components
correspond to the Maxwell fields).  Without additional
constraints,  the particles and fields are not confined to their mass
shells. We show that in the
mass-shell limit, the generalized Lorentz force obtained by means of 
the retarded Green's functions for the five dimensional field equations 
provides the classical Abraham-Lorentz-Dirac radiation reaction terms
(with renormalized mass and charge).  We also
obtain general coupled equations for the orbit and the off-shell
 dynamical mass during the evolution.  The theory does not admit
 radiation if the particle remains identically on-shell. The  structure of 
the equations implies that
 mass-shell deviation is bounded when the external field is removed.

\vfill
\break
\bigskip
\par  Gupta and Padmanabhan$^4$ have shown that the motion of a charged
 particle in an electromagnetic field 
  can be described in the inertial frame of the particle with a time
  varying non-trivial background metric. Using the general covariant
  form of the Maxwell equations and transforming back to the inertial
  frame of the laboratory, they obtained the
Abraham-Lorentz-Dirac radiation reaction term as a consequence of this
geometrical picture.  This result demonstrates that the description of
  the motion of a charged particle in acceleration must include the
 radiation terms of the Abraham-Lorentz-Dirac equation$^5$. 
 \par Alternatively, one can develop the mechanics in a flat space of
higher dimension, an approach that we shall take.
 We shall work with the manifestly covariant
mechanics of Stueckelberg$^6$, which provides a description of
dynamical systems under the influence of forces (which may be
represented in terms of potentials or gauge fields) in a framework which is
Lorentz covariant. This theory admits, on a classical level,
  deviations from the particle's mass shell during interaction, as in
  quantum field theory. A similar
  approach was used by Mendon\c ca and Oliveira
e Silva$^7$, who studied the motion of a relativistically kicked
oscillator in the $E,\, t$ plane using what they called a ``super
Hamiltonian.''  One can, in fact derive the relativistic Lorentz force 
$$ m {\ddot x}^\mu = F^\mu\,_\nu {\dot x}^\nu  \eqno(1)$$  
from such a Hamiltonian.
\par  Consider the Hamiltonian$^{5,6}$(we take
$c=1$ henceforth) 
$$ K = {(p^\mu - e A^\mu(x))(p_\mu - e A_\mu(x)) \over 2M }\eqno(2),$$
where $x \equiv x^\mu $. The Hamilton equations (generalized to the
 four-dimensional symplectic mechanics$^6$) are 
$$ \eqalign{ {dx^\mu \over d\tau} &= {\partial K \over \partial p_\mu}= 
{p^\mu - eA^\mu(x) \over M } \cr
   {dp^\mu \over d\tau } &= - {\partial K \over \partial x_\mu} = 
e {\partial A^\lambda(x) \over \partial x_\mu}{ p^\lambda - e A^\lambda (x)
\over M}, \cr}\eqno(3)$$
where $\tau$ is the absolute (universal) invariant time parametrizing the path
 of the particle in spacetime$^6$.
 Computing $  {dp^\mu \over d\tau }$ from the first of these, one
 finds Eq. $(1)$.  It moreover follows from the first of Eqs.$ (3)$ that
$$ {dx^\mu \over d\tau}{dx_\mu \over d\tau} =  {(p^\mu - e
A^\mu(x))(p_\mu - e A_\mu(x)) \over M^2 }; \eqno(4)$$
this quantity is absolutely conserved, since $K$ does not depend
explicitly on $\tau$. 
It follows, since the square of the proper time $ds^2= - dx^\mu dx_\mu$, that
$ds$ is proportional to $d\tau$, independently of the acceleration of
the particle.
   The numerator of $(4)$ is the mass-squared  of
the particle; we infer that this result is associated with the
restriction of the particle to a sharp mass shell.
\par  Taking into account
 full $U(1)$ gauge invariance, the Stueckelberg-Schr\"odinger \hfil\break
equation$^6$ (including a compensation field for the
$\tau$-derivative) is
$$ \bigl(i{\partial \over \partial \tau} + e_0 a_5 \bigr) \psi_\tau (x) = 
  {(p^\mu - e_0a^\mu(x,\tau))(p_\mu - e_0 a_\mu(x,\tau)) \over 2M
}\psi_\tau(x), \eqno(5)$$
where the gauge fields may depend on $\tau$ and $e_0$ is a
dimensionless coupling.
The corresponding classical Hamiltonian then has the form
 $$ K=  {(p^\mu - e_0a^\mu(x,\tau))(p_\mu - e_0 a_\mu(x,\tau)) \over 2M
}- e_0 a_5(x,\tau) . \eqno(6)$$
The equations of motion for the field variables are given (for both the
 classical and quantum theories) by$^8$
$$ \lambda \partial_\alpha f^{\beta \alpha}(x,\tau) = e_0
j^\beta(x,\tau), \eqno(7)$$
where $\alpha,\, \beta = 0,1,2,3,5$, the last corresponding to the
$\tau$ index,  and $\lambda$, of dimension $\ell^{-1}$, 
 is a factor on the terms $f^{\alpha
\beta} f_{\alpha \beta}$ in the Lagrangian associated with $(6)$
(including degrees of freedom of the fields), required by dimensionality,
as we shall see below.  The field strengths are 
$$ f^{\alpha \beta} = \partial^\alpha a^\beta - \partial^\beta
a^\alpha, \eqno(8)$$
and the current satisfies the conservation law$^{8,10}$
$$ \partial_\alpha j^\alpha(x,\tau) = 0. \eqno(9)$$
Writing out $(9)$ explicitly ($j^5 \equiv \rho$,
the density of events in spacetime), 
$$ \partial_5 \rho + \partial_\mu j^\mu =0; \eqno(10)$$
integrating over $\tau$ on $(-\infty, \infty)$, and assuming that
$j^5(x,\tau)$ vanishes$^8$ at $\vert \tau \vert \rightarrow \infty$, 
one finds that 
$$\partial_\mu J^\mu(x) = 0, $$
where (for some dimensionless $\eta$) 
$$ J^\mu(x) = \eta \int_{-\infty}^\infty \, d\tau \, j^\mu(x,\tau).
\eqno(11)$$  
We identify this $J^\mu(x)$ with the Maxwell conserved current.
In ref. 9, for example, this expression occurs with
    $$ j^\mu (x,\tau) = {\dot x}^\mu(\tau) \delta^4(x-x(\tau)),
\eqno(12)$$
and $\tau$ is identified with the proper time of the particle
 (an identification which can be
made for the motion of a free particle).
 
\par  Integrating the $\mu$-components of
Eq. $(7)$ over $\tau$ (assuming $f^{\mu 5} (x, \tau) \rightarrow 0$
 for $\tau \rightarrow \pm \infty$), we obtain the Maxwell
 equations with the Maxwell charge $e= e_0/\eta$ and
 the Maxwell fields  given by 
$$ A^\mu(x) = \lambda \int_{-\infty}^\infty a^\mu(x,\tau) \,d\tau. \eqno(13)$$
 The Hamiltonian of
Stueckelberg$^6$ and Mondon\c ca and Oliveira e Silva$^8$ can be
recovered in the limit of the zero mode of the fields
$$ a^\mu(x,\tau) = \int \,ds {\hat a}^\mu (x,s)
e^{-is\tau}. \eqno(14)$$
In the zero mode limit, when the Fourier transform of the fields have
support only in the neighborhood $\Delta s$ of $s=0$, the vector
potential takes on the form
 $  a^\mu(x,\tau) \sim \Delta s {\hat a}^\mu(x,0)
 =(\Delta s / 2\pi \lambda) A^\mu(x)$, and we
identify $e= (\Delta s/2\pi \lambda) e_0$. The zero mode therefore
emerges when the inverse correlation length of the field satisfies the
relation $ \eta \Delta s = 2\pi \lambda $. We remark that in this
limit, the fifth equation obtained from $(7)$ decouples; the zero mode
of the $\tau$ derivative of $a^\mu(x,\tau)$ vanishes. If the parameter
$\lambda$ is independent of the dynamical structure of the fields, then
 the effective
width of ${\hat a}^\mu(x,s)$, when it is well-defined, affects the
value of the charge $e$, as well as the relation between the effective
Maxwell current and the microscopic current $j^\mu$.This effect, occurring
when a Maxwell type theory is a good approximation, can be understood
as a classical analog of charge renormalization, where the
effective charge is a function of momentum transfer.

\par Again, writing the Hamilton equations for the Hamiltonian $(6)$,
we find the generalized Lorentz force$^{10}$ 
$$ M {\ddot x}^\mu = e_0 f^\mu\,_\nu {\dot x}^\nu +
f^\mu\,_5 = e_0 \bigl({f_{self}}^\mu\,_\nu {\dot x}^\nu +
{f_{self}}^\mu\,_5 +
{f_{ext}}^\mu\,_\nu {\dot x}^\nu  +{f_{ext}}^\mu\,_5 \bigr). \eqno(15)$$
Multiplying this equation by ${\dot x}_\mu $, one obtains
$$ M{\dot x}_\mu {\ddot x}^\mu  = e_0{\dot x}_\mu f^\mu\,_5=e_0\bigl({\dot x}_\mu {f_{self}}^\mu\,_5+{\dot x}_\mu {f_{ext}}^\mu\,_5\bigr); \eqno(16)$$
this equation therefore does not necessarily lead to the trivial
relation between $ds$ and $d\tau$ discussed above in
connection with Eq. $(4)$. The $f^\mu \,_5$ term has the effect of moving
the particle off-shell. 
\par In the following we use the Green's functions for $(7)$ to calculate
the radiation reaction force directly, as, for example, in the derivation
of Sokolov and Ternov$^{11}$.
  In the limit for which the particle stays on its mass shell during the 
interaction, we show that this formula reduces to the known
Abraham-Lorentz-Dirac formula$^{9,12}$ 
for the Maxwell self-interaction problem. We furthermore show that the
deviation from mass shell is stable. We shall use the retarded Green's
function and treat divergences by renormalization of charge and the
mass parameter $M$.     
\par Choosing the generalized Lorentz gauge $\partial_\alpha a^\alpha
=0$, Eq. $(7)$ becomes
$$ \lambda \partial_\alpha \partial^\alpha a^\beta(x,\tau)= (\sigma
\partial_\tau^2 - \partial_t^2 + \bigtriangledown^2) a^\beta =
-e_0 j^\beta(x,\tau),  \eqno(17)$$
where $\sigma = \pm1$ corresponds to the possible choices of metric
for the symmetry ${\rm O}(4,1)$ or
${\rm O}(3,2)$ of the homogeneous field equations.
\par The Green's functions for Eq. $(17)$ can be constructed from the
inverse Fourier transform
$$ G(x,\tau) = { 1 \over (2\pi)^5} \int \, d^4k d\kappa {e^{i(k^\mu
x_\mu + \sigma \kappa \tau)} \over k_\mu k^\mu   + \sigma
\kappa^2}. \eqno(18)$$
Integrating this expression over all $\tau$ gives the Green's function
for the standard Maxwell field.  Assuming that the radiation
reaction acts causally in $\tau$, we shall restrict our attention
here to the $\tau$-retarded Green's function.  In his calculation of
the radiation corrections to the Lorentz force,  Dirac$^{12}$
used the difference between advanced and retarded  Green's
functions in order to cancel the singularities that they
contain.  One can, alternatively$^{11}$,
use the retarded Green's function and ``renormalize'' the mass in
order to eliminate the singularity.  In our analysis, we follow the
latter procedure.
\par The $\tau$- retarded Green's function is given by multiplying
the principal part of the integral $(18)$ by $\theta(\tau)$.
  Carrying out the integrations
(on a complex contour in $\kappa$; we consider the case 
 $\sigma = +1$ in the following), one finds (this Green's function
 differs from that used in ref. 13,
constructed on a complex contour in $k^0$)
$$ G(x,\tau) = {2 \theta (\tau) \over (2\pi)^3}\cases{{\tan^{-1}
\bigl({\sqrt{ -x^2-\tau^2} \over \tau} \bigr) \over (-x^2 -\tau^2)^{3
\over 2}} - {\tau \over x^2(x^2 + \tau^2)}  &$x^2 + \tau^2 < 0$; \cr
{ 1 \over 2}{1 \over (\tau^2 + x^2)^{3 \over 2}}  \ln\bigl\vert {\tau -
 \sqrt{\tau^2 + x^2} \over  \tau +
 \sqrt{\tau^2 + x^2}} \bigr\vert  - {\tau \over
x^2(\tau^2 + x^2)} &$x^2 + \tau^2 > 0$. \cr} \eqno(19)$$
\par With the help of this Green's function, the solutions of
 Eq. $(17)$ for the self-fields can be written,
$$ \eqalign{{a_{self}}^\mu (x, \tau) &= {e_0 \over \lambda}\int\,d^4x'd\tau' G(x-x',\tau-\tau'){\dot
x}^\mu(\tau') \delta^4(x'-x(\tau'))\cr
&= {e_0 \over \lambda}\int \, d\tau' {\dot x}^\mu (\tau')G(x - x(\tau'),\tau-\tau') \cr
{a_{self}}^5(x,\tau) &=  {e_0 \over \lambda}\int\,d^4x'd\tau' G(x-x',\tau-\tau')
 \delta^4(x'-x(\tau'))\cr
&= {e_0 \over \lambda}\int \, d\tau' G(x - x(\tau'),\tau-\tau') \cr}
 \eqno(20)$$
where we have used (12) (along with
$j^5(x,\tau)=\delta^4(x-x(\tau)))$.
We have written this Green's function as a scalar, acting in the same
way on all five components of the source $j^\alpha$; to assure that
the resulting field is in Lorentz gauge, however, it should be written
as a five by five matrix, with the factor
 $\delta^\alpha_\beta - k^\alpha k_\beta /k^2$ ($k_5=\kappa$)
included in the integrand.  Since we are computing only the gauge invariant field strengths here, this extra term will not influence any of the results. 
\par From $(8)$ and $(15)$, it then follows that the generalized
Lorentz force for the self-action (the force of the fields generated by
the world line on a point $x^\mu(\tau)$ of the trajectory) is
$$ \eqalign{M{\ddot x}^\mu &= {e_0^2 \over \lambda} \int\, d\tau' ({\dot x}^\nu(\tau)
{\dot x}_\nu(\tau')\partial^\mu - {\dot x}^\nu(\tau){\dot
x}^\mu(\tau')
\partial_\nu) G(x-x(\tau'))\vert_{x=x(\tau)} \cr
 &+ {e_0^2 \over \lambda}\int \,d\tau' (\partial^\mu - {\dot x}^\mu(\tau')
\partial_\tau) G(x-x(\tau'))\vert_{x=x(\tau)}\cr &+e_0 \bigl({f_{ext}}^\mu\,_\nu {\dot x}^\nu +
 {f_{ext}}^\mu\,_5 \bigr)\cr} \eqno(21)$$
\par We define $ u \equiv (x_\mu(\tau) - x_\mu(\tau'))(x^\mu(\tau)
-x^\mu(\tau'))$,
 so that
 $$ \partial_\mu = 2(x_\mu(\tau) - x_\mu(\tau')){\partial \over \partial
u}.\eqno(22)$$
Eq. (21) then becomes
$$ \eqalign{M{\ddot x}^\mu&=2{e_0^2 \over \lambda} \int\, d\tau' \{{\dot x}^\nu(\tau)
{\dot x}_\nu(\tau')(x^\mu(\tau)-x^\mu(\tau'))\cr
& - {\dot x}^\nu(\tau){\dot
x}^\mu(\tau')
(x_\nu(\tau)-x_\nu(\tau'))\}{\partial \over \partial u}
G(x-x(\tau'),\tau-\tau')
\vert_{x=x(\tau)}) \cr
 &+ {e_0^2 \over \lambda}\int \,d\tau' \{2(x^\mu(\tau)-x^\mu(\tau')){\partial
\over \partial u}
 - {\dot x}^\mu(\tau')
\partial_\tau\} G(x-x(\tau'),\tau-\tau')\vert_{x=x(\tau)}).\cr &+e_0 \bigl({f_{ext}}^\mu\,_\nu {\dot x}^\nu +{f_{ext}}^\mu\,_5 \bigr)\cr} \eqno(23)$$
\par We now expand the integrands in Taylor series around the most singular
point $\tau=\tau'$. In this neighborhood, keeping the lowest order terms in $\tau''=\tau-\tau'$, the variable $u$ reduces to $u\cong {\dot x}^\mu{\dot x}_\mu{\tau''}^2$. We shall also use the following definition;
$$\varepsilon \equiv 1+{\dot x}^\mu{\dot x}_\mu, \eqno(24)$$
a quantity that vanishes on the mass
shell of the particle (as we have pointed out above).
In this case the derivatives of $(19)$ take the form
$$\eqalign{{{\partial G}\over{\partial u}} &\cong{{\theta(\tau'')f_1 (\epsilon)}\over {(2\pi)^3{\tau''}^5}} \cr
{{\partial G}\over{\partial\tau''}} &\cong{{\theta(\tau'')f_2 (\epsilon)}\over {(2\pi)^3{\tau''}^4}}+ {{\delta(\tau'')f_3 (\epsilon)}\over {(2\pi)^3{\tau''}^3}}\cr} \eqno(25) $$
where we have used the following definitions:

\noindent $\varepsilon<0$:
$$\eqalign{f_1(\varepsilon) &= {3\tan^{-1}(\sqrt{-\varepsilon}) \over{(-\varepsilon)}^{5 \over 2}}-{3 \over \varepsilon^2(1-\varepsilon)}+{2 \over \varepsilon(1-\varepsilon)^2} \cr
f_2(\varepsilon) &= {3\tan^{-1}(\sqrt{-\varepsilon}) \over{(-\varepsilon)}^{5 \over 2}}-{1\over \varepsilon^2}-{2 - \varepsilon  \over \varepsilon^2(1-\varepsilon)}\cr
f_3(\varepsilon) &= {\tan^{-1}(\sqrt{-\varepsilon})
\over{(-\varepsilon)}^{3 \over 2}}+{1\over
\varepsilon(1-\varepsilon)}\cr}\eqno(26a)$$

\noindent $\varepsilon>0$:
$$\eqalign{f_1(\varepsilon) &= {{3\over
2}\ln\bigl\vert{1+\sqrt{\varepsilon}
 \over 1-\sqrt{\varepsilon}}\bigr\vert \over{(\varepsilon)}^{5 \over 2}}-{3 \over \varepsilon^2(1-\varepsilon)}+{2 \over \varepsilon(1-\varepsilon)^2} \cr
f_2(\varepsilon) &= {{3\over 2}\ln\bigl\vert{1+\sqrt{\varepsilon} \over 1-\sqrt{\varepsilon}}\bigr\vert \over{(\varepsilon)}^{5 \over 2}}-{1\over \varepsilon^2}-{2 - \varepsilon  \over \varepsilon^2(1-\varepsilon)}\cr
f_3(\varepsilon) &=- {{1\over 2}\ln\bigl\vert{1+\sqrt{\varepsilon}
\over 1-\sqrt{\varepsilon}}\bigr\vert \over{(\varepsilon)}^{3 \over
2}}+{1\over \varepsilon(1-\varepsilon)}\cr}\eqno(26b)$$
For either sign of $\varepsilon$, when $\varepsilon \sim 0$, 
$$\eqalign{ f_1(\varepsilon) &\sim {8 \over 5} +{24 \over 7}
\varepsilon, \cr f_2(\varepsilon) &\sim -{2 \over 5} - {4 \over 7}
\varepsilon, \cr f_3(\varepsilon) &\sim {2 \over 3} + {4 \over 5}
\varepsilon \cr}\eqno(26c) $$
One sees that the derivatives in (25) have no singularity in
$\varepsilon$ at $\varepsilon=0$.

\par From $(8)$ and $(20)$, we have
$$\eqalign{ &{f_{self}}^\mu \,_5 (x(\tau), \tau) =\cr & e \int \,d\tau'\{2(x^\mu(\tau)-x^\mu(\tau')){\partial\over \partial u}- {\dot x}^\mu(\tau')
\partial_\tau\} G(x-x(\tau'),\tau-\tau')\vert_{x=x(\tau)},\cr} \eqno(27)$$
  We see (from $(25)$) that the main contributions to the integrals
 come from small
$\tau''$.  We may therefore expand $x^\mu(\tau) -
x^\mu(\tau')$ and ${\dot x}^\mu(\tau) -
{\dot x}^\mu(\tau')$ in $(27)$ in power series in $\tau''$, and write
the integrals formally with infinite limits. 
\par Substituting $(27)$ into $(16)$, we obtain (note that $x^\mu$ and
its derivatives are evaluated at the point $\tau$, and are not subject to
the $\tau''$ integration), after integrating by parts using $\delta(\tau'')=
{\partial\over \partial \tau''}\, \theta(\tau'')$,
$$ \eqalign{M {\dot x}_\nu {\ddot x}^\nu =  {2e_0^2 \over \lambda
(2\pi)^3} \int_{-\infty}^\infty d\tau''\bigl\{{(f_1-f_2-3f_3) \over \tau''^4}{\dot x}_\nu{\dot x}^\nu- &{({1 \over 2}f_1-f_2-2f_3)\over\tau''^3}
{\dot x}_\nu {\ddot x}^\nu \cr  +  {({1 \over 6}f_1-{1 \over 2}f_2-{1 \over 2}f_3)\over \tau''^2}
{\dot x}_\nu {\mathop{x}^{...}}^\nu 
\bigr\} \theta (\tau'')+e_0{\dot x}_\mu{f_{ext}}^\mu\,_5 . \cr} \eqno(28)$$
The integrals are divergent at the lower bound $\tau''=0$ imposed by
the $\theta$-function; we therefore take these integrals to a cut-off
$\mu >0$.  Eq.$(28)$ then becomes
$$ \eqalign{{M \over 2} {\dot \varepsilon} =  {2e_0^2 \over \lambda
(2\pi)^3} \bigl\{{(f_1-f_2-3f_3) \over 3\mu^3}(\varepsilon - 1)- &{({1
\over 2}f_1-f_2-2f_3)\over 4\mu^2}{\dot \varepsilon} \cr  +  {({1 \over 6}f_1-{1 \over 2}f_2-{1 \over 2}f_3)\over \mu}
{\dot x}_\nu {\mathop{x}^{...}}^\nu 
\bigr\}  + e_0{\dot x}_\mu{f_{ext}}^\mu\,_5 . \cr} \eqno(29)$$
 
\par Following a similar procedure, we obtain from $(23)$

$$ \eqalign{M{\ddot x}^\mu &= {2e_0^2 \over \lambda (2\pi)^3}
 \bigl\{ -{f_1\over 4 \mu^2}
 ((1-\varepsilon){\ddot x}^\mu + {1 \over 2}{\dot\varepsilon} {\dot
x}^\mu)  + {f_1\over 3 \mu} ({\dot x}_\nu{\mathop{x}^{...}}^\nu {\dot
x}^\mu+(1-\varepsilon){\mathop{x}^{...}}^\mu) \cr &+{(f_1-f_2-3f_3)
\over 3\mu^3}{\dot x}^\mu- {({1 \over 2}f_1-f_2-2f_3)\over
2\mu^2}{\ddot x}^\mu +
 {({1 \over 6}f_1-{1 \over 2}f_2-{1 \over 2}f_3)\over
 \mu}{\mathop{x}^{...}}^\mu \bigr\}
 \cr &+e_0 \bigl({f_{ext}}^\mu\,_\nu {\dot x}^\nu +{f_{ext}}^\mu\,_5  \bigr).\cr} \eqno(30)$$
Using $(29)$ to substitute for the coefficient of the $ {1 \over
\mu^3} $ term in $(30)$ , we obtain (for $\varepsilon \neq 1$)
 $$\eqalign{M(\varepsilon) {\ddot x}^\mu &=-{1 \over 2}{M(\varepsilon) \over
(1-\varepsilon)} {\dot \varepsilon} {\dot x}^\mu+ {2e_0^2 \over
\lambda (2\pi)^3 \mu}
  F(\varepsilon) \bigl\{{\mathop{x}^{...}}^\mu + { 1\over
 (1-\varepsilon)} {\dot x}_\nu{\mathop{x}^{...}}^\nu {\dot x}^\mu \bigr\}\cr &+{e_0{\dot
 x}^\mu{\dot x}_\nu{f_{ext}}^\nu\,_5 \over 1-\varepsilon}
+e_0{f_{ext}}^\mu\,_\nu {\dot x}^\nu +e_0{f_{ext}}^\mu\,_5  , \cr}  \eqno(31)$$
where
$$ F(\varepsilon)={f_1 \over 3}(1-\varepsilon)+({1 \over 6}f_1-{1
\over 2}f_2-{1 \over 2}f_3).  \eqno(32)$$
\par Here, the coefficients of ${\ddot x}^\mu$ have been grouped into a 
renormalized (off-shell) mass term, defined (as in
the procedure of Sokolov and Ternov$^{11}$) as
$$ M(\varepsilon) = M + {e^2 \over 2\mu}\bigl[{f_1(1-\varepsilon) \over 2}+{1 \over 2}f_1-f_2-2f_3\bigr]
\eqno(33)$$
where, as we shall see below,
$$ e^2 =  {2e_0^2 \over \lambda (2\pi)^3\mu}, \eqno(34)$$
can be identified with the Maxwell charge by studying the on-shell limit. 

\par  We now obtain, from $(31)$,
$$\eqalign{ M(\varepsilon) {\ddot x}^\mu &= -{1 \over 2} {M(\varepsilon)\over
 {1-\varepsilon}}{\dot \varepsilon} {\dot x}^\mu + F(\varepsilon)
 e^2\bigl\{ {\mathop{x}^{...}}^\mu+ {1 \over 1-\varepsilon}{\dot x}_\nu
{\mathop{x}^{...}}^\nu {\dot x}^\mu \bigr\}  \cr &+ e_0 {f_{ext}}^\mu_\nu
{\dot x}^\nu + e_0\Bigl({{\dot x}^\mu
{\dot x}_\nu \over 1-\varepsilon}+\delta^\mu_\nu \Bigr){f_{ext}}^\nu_5. \cr} \eqno(35)$$
We remark that when one multiplies this equation by ${\dot x}_\mu$,
it becomes an identity (all of the terms except for $
e_0 {f_{ext}}^\mu_ \nu{\dot x}^\nu$ may be grouped to be proportional
to $\Bigl( {{\dot x}^\mu{\dot x}_\nu \over
1-\varepsilon}+\delta^\mu_\nu
 \Bigr)$); one must use Eq. $(29)$ to compute the
off-shell mass shift $\varepsilon$ corresponding to the longitudinal
degree of freedom in the direction of the four velocity of the
particle.  
Eq. $(35)$ determines the motion orthogonal to the four velocity.
Equations $(29)$ and $(35)$ are the fundamental dynamical equations
governing the off-shell orbit.
\par We now show that the standard relativistic Lorentz force, with
radiation corrections, can be obtained from these equations when
$\mu{\dot \varepsilon} << \varepsilon <<1$ and ${\ddot \varepsilon}$
and $f_{ext}^\mu\,_5$ are
small.  In this case, Eq. $(29)$ becomes
$$ \bigl(M- {1\over 15 \mu}\bigr) {{\dot\varepsilon} \over 2} \cong e^2 \Bigl\{ -{8\varepsilon \over 15\mu^2 } +
{2 \over 15} {\dot x}_\nu {\mathop{x}^{...}}^\nu \Bigr\} \eqno(36)
$$
The left hand side can be neglected if 
$$\Bigl[ M/({e^2 \over \mu}) \Bigr] (\mu {\dot \varepsilon})<<
\varepsilon. \eqno(37)$$
We shall see below that we must have $0.68\, e^2/\mu <M$ for stability of
$\varepsilon$, but if $e^2/\mu$ is not too small, the inequality
$(37)$ is consistent with our assumed inequalities, and it then
follows that
$$ 4\varepsilon / \mu^2 \cong  {\dot x}_\nu {\mathop{x}^{...}}^\nu.
\eqno(38)$$ 
If, furthermore, ${\ddot \varepsilon}$ is small, then
$${\dot x}_\mu {\mathop{x}^{...}}^\mu
= {\ddot \varepsilon} - {\ddot x}_\mu{\ddot x}^\mu \cong -{\ddot
x}_\mu{\ddot x}^\mu,\eqno(39)$$
the known expression associated with radiation.  Since $\varepsilon/\mu^2$ may be
appreciable even if $\varepsilon$ is small, the inequalitites we have
assumed can admit a significant contribution of this type.
Under these conditions equation $(34)$ becomes,
 $$M_{ren} {\ddot x}^\mu =   {2 \over 3}e^2 \{ {\mathop{x}^{...}}^\mu- {\ddot
x}_\nu{\ddot x}^\nu {\dot x}^\mu \}+e_0{f_{ext}}^\mu\,_\nu {\dot
x}^\nu, \eqno(40)$$
where $M_{ren} = M(\varepsilon)|_{\varepsilon = 0} = M + e^2/3\mu$.
\par This result is of the form of the standard relativistic Lorentz
force with radiation reaction.$^{9,11,12,14}$
\par We now study the stability of the variations of the off-shell
parameter $\varepsilon$ when the external field  is removed. 
First, we construct an  equation of motion for $\varepsilon$.  We
define the functions
$$\eqalign{F_1(\varepsilon) &= { 1 \over 3\mu^2}(\varepsilon - 1)
(f_1-f_2-3f_3) \cr
F_2(\varepsilon) &= {1 \over 4\mu}({1 \over 2} f_1 - f_2 -2f_3) \cr
F_3(\varepsilon) &= {1 \over 6}f_1 -{1\over 2}f_2 -{1 \over 2}f_3 \cr}
 \eqno(41)$$
 so equation $(29)$, in the absence of external
fields, becomes:
$${M\over 2}{\dot \varepsilon}=e^2 \bigl\{F_1(\varepsilon)+F_2(\varepsilon){\dot \varepsilon}+F_3(\varepsilon){\dot x_\mu  {\mathop{x}^{...}}^\mu} \bigr\}. \eqno(42)$$ 
 Solving for the explicit $x$ derivatives in $(42)$ and differentiating with
respect to $\tau$, one obtains
$$\eqalign{{\dot x_\mu  {\mathop{x}^{....}}^\mu}+{\ddot x_\mu
{\mathop{x}^{...}}^\mu}=&{1 \over F_3}\bigl\{F'_2{\dot
\varepsilon}^2+{\ddot \varepsilon}\bigl({M\over 2 e^2}+F_2 \bigr)-F'_1
{\dot \varepsilon} \bigr\} \cr &-{F'_3  \over {F_3}^2}\bigl\{F_2+ {M
\over 2 e^2}{\dot \varepsilon}-F_1 \bigr\}{\dot \varepsilon}\equiv
H.\cr} 
 \eqno(43)$$
Together with $${\dot x_\mu  {\mathop{x}^{....}}^\mu}+3{\ddot x_\mu  {\mathop{x}^{...}}^\mu}= {1\over 2}{\mathop{\varepsilon}^{...}}$$
one finds, from $(43)$, 
$$ {\ddot x_\mu  {\mathop{x}^{...}}^\mu}={1\over 4}{\mathop{\varepsilon}^{...}}-{1\over2}H(\varepsilon,{\dot \varepsilon}, {\ddot \varepsilon}) \eqno(44)$$
Multiplying Eq.$(35)$ by ${\ddot x}_\mu$ (with no external fields) and
using $(42)$ and $(44)$, we obtain
$$  {\mathop{\varepsilon}^{...}}-A(\varepsilon){\ddot \varepsilon}+
B(\varepsilon){\dot \varepsilon}^2 + C(\varepsilon) {\dot
\varepsilon}-D(\varepsilon)=0, \eqno(45)$$
where
$$\eqalign{ A(\varepsilon) &= {2 \over F_3}\bigl( {M \over
2e^2} + F_2 \bigr) + {2 M(\varepsilon) \over e^2 F(\varepsilon)},\cr
B&= {2F_3' \over F_3^2}\bigl( {M\over 2e^2} + F_2 \bigr) - {2F_2'
\over F_3} + {2 \over {1-\varepsilon}} {1 \over F_3} \bigl( {M\over
2e^2} + F_2 \bigr)-{M(\varepsilon)\over e^2F(\varepsilon)} {1 \over
1-\varepsilon}, \cr
C&= {4 M(\varepsilon) \over e^2 F(\varepsilon)} {1 \over F_3}\bigl(
{M\over 2e^2} + F_2 \bigr) - {2  \over F_3^2}F_1 F_3'\cr
&-{2F_1 \over (1-\varepsilon)F_3} + {2 \over F_3}F_1' ,\cr
D &= {4 M(\varepsilon) \over e^2 F(\varepsilon)}{F_1 \over F_3}. \cr}
\eqno(46)$$   
\par We first study the possibilty of having a solution of the form
$\varepsilon\equiv \varepsilon_0$, a constant.
In this case ${\ddot \varepsilon}=0$ implies,
 $$ \dot x^\mu  {\mathop{x}^{...}}_\mu=-{\ddot x}_\mu {\ddot x}^\mu.$$
Since all the derivatives of $\varepsilon$ are zero we also find from $(44)$,
$$  {\ddot x_\mu  {\mathop{x}^{...}}^\mu}=0$$.
Multiplying eq.$(35)$ by ${\ddot x_\mu}$ and substituting these last two results we get 
$${\ddot x_\mu} {\ddot x^\mu}= {\dot x^\mu}  {\mathop{x}^{...}}_\mu=0$$ 
From $(42)$ we find then that
 $${\dot x}_\mu {\mathop{x}^{...}}^\mu=-{F_1 \over F_3}=0.$$
From $(41)$ and $(26b)$, one sees that
this equation can be satisfied only if $\varepsilon=0$ ($F_1 = 0$)
or $\varepsilon=1$ ($F_3 = \infty$).

\par Since ${\dot \varepsilon}=0$ we find that ${\dot t}^2{\ddot
t}^2={\vert{\dot {\bf x}}\vert}^2 {\vert{\ddot {\bf x}}\vert}^2 cos^2{\theta}$.  Together with ${\ddot x}_\mu{\ddot x}^\mu={\ddot t}^2-\vert {\ddot
{\bf x}}^2 \vert=0$ this implies
$${\dot t}^2{\ddot t}^2={\vert{\dot x}\vert}^2 {\ddot t}^2 \cos^2{\theta}$$ 
The solution ${\ddot t}=\vert{\ddot x}\vert=0$ implies ${\dot x}_\mu={\rm const}$ .
The other solution ${\dot t}^2={\vert{\dot x}\vert}^2 \cos^2{\theta}$
implies that $\vert {\dot{\bf x}}\vert^2 (1- \cos^2{\theta}) =
\varepsilon -1$; since the left hand side is positive, $\varepsilon$
cannot be zero, and the only possibility for a constant solution is
 then $\varepsilon =1$, motion on the light cone. We shall show below that the trajectory cannot
reach this bounday. 
\par  The mass shell condition $\varepsilon
 =0$, in the theoretical framework we have given here, implies that
 the particle motion must be with constant velocity, and that no
 radiation (${\ddot x_\mu} {\ddot x^\mu} =0$)
 is possible, {\it i.e.}, in order to radiate, the particle must be
 off-shell.  This result is also true in the presence of an external field. In particular, it follows from Eq. $(29)$ that for $\varepsilon \equiv 0$,
$$\eqalign{ -{2 \over 15\mu} {\dot x^\mu}  {\mathop{x}^{...}}_\mu
&\equiv {2 \over 15\mu}{\ddot x}_\mu {\ddot x}^\mu \cr &= 
  e_0{\dot x}_\mu{f_{ext}}^\mu\,_5 .\cr} \eqno(47)$$
From Eq.$(15)$, however, it follows (in case ${\dot \varepsilon}=0$)
that ${\dot x}_\mu{f_{ext}}^\mu\,_5 = -{\dot
x}_\mu{f_{self}}^\mu\,_5$, so that the nonvanishing value of ${\ddot
x}_\mu {\ddot x}^\mu$ corresponds only to a self-acting field
${f_{self}}^\mu\,_5$ (driven by ${f_{ext}}^\mu\,_5$), and not to radiation.
\par We  now show that, in general, $\varepsilon$ is bounded when the external fields are turned off.  For the case $\varepsilon<0$ the function $F_3$ is zero at $\varepsilon=-0.735$. In this case eq.$(42)$ becomes
$${\dot \varepsilon}(-0.735)={F_1(-0.735) \over {M \over
2e^2}+F_2(-0.735)}. \eqno (48)$$
If ${\dot\varepsilon} >0$ at this value of $\varepsilon$, then
$\varepsilon$ cannot cross this boundary.  Since 
 $F_1(-0.735)={.624 \over \mu^2}$ , $F_2(-0.375)=-{.259 \over \mu}$,
this condition implies that
$$  \mu > 0.68 {e^2 \over M}.$$
Setting $M, e$ equal to the electron mass (the lowest mass charged
particle) and charge one finds that $\mu>10^{-23}$sec, a cut-off of
reasonable size for a classical theory.
\par We now show that $\epsilon$ is bounded from above by unity. The
full classical Hamiltonian, obtained by adding the contribution of the fields
to the expression on the right hand side of $(6)$, is a conserved
quantity. In the absence of external fields, all the field quantities
are related to the source particle through the Green's functions. In
the absence of external fields, as the particle motion approaches the
light cone, there are infinite contributions
arising from the fields evaluated on the particle trajectory. In this
case, it follows from $(4)$ that ${(p^\mu - e_0a^\mu(x,\tau))(p_\mu -
 e_0 a_\mu(x,\tau)) \over 2M} =0$. The $a_5$ self-field term is less
singular than the $f_{\mu\nu}f^{\mu\nu}$ and $f_{\mu 5}f^{\mu}\,_5$
terms, which involve derivatives of the Green's functions, as in
$(25)$, squared.  As seen from $(26b)$, the most singular contribution
arises from $f_1\,^2$. Since the total Hamiltonian $K$ is conserved,
the coefficient of this singularity must vanish.  The coefficients
involve just $\dot \varepsilon$ (and its square) and $\ddot
\varepsilon$; one finds a simple nonlinear differential equation for
which only ${\dot \varepsilon}=0$ can be a
solution. It follows that the conservation law restricts the
evolution of $\varepsilon$ to values less than unity, {\it i.e.}
the particle trajectory cannot pass through the light cone.
\par This bound manifests itself in the structure of the differential
equation $(45)$ for $\varepsilon$.  In the limit that $\varepsilon
\rightarrow 1$, the coefficients $A,\, B,\, C,\, D$ are all finite;
however the behavior of the linearized solution depends on the
derivates of these coefficients, and, in this limit, $B'$ is singular,
driving the solution away from the light cone.
 \par  Numerical studies are under way to follow the motion of this
 highly nonlinear system both in the presence and absence of external fields.

\bigskip
\noindent
{\it Acknowledgements:\/} We wish to thank Y. Ashkenazy,
 J. Bekenstein, C. Piron and F. Rohrlich and S.L. Adler for helpful
 discussions, and Z. Schuss and J. Schiff for explaining some things about
 non-linear equations to us.
 One of us (L.H.) also wishes to thank S.L. Adler for his hospitality at
 the Institute for Advanced Study where much of this work was done.
\bigskip
\noindent
{\it References}
\frenchspacing
\smallskip
\item{1.} G.M. Zaslavskii, M.Yu. Zakharov, R.Z. Sagdeev,
D.A. Usikov, and A.A. Chernikov, Zh. Eksp. Teor. Fiz {\bf 91}, 500 (1986)
[Sov. Phys. JEPT {\bf 64}, 294 (1986)].
\item{2.} D.W. Longcope and R.N. Sudan, Phys. Rev. Lett. {\bf 59},
1500 (1987).
\item{3.} H. Karimabadi and V. Angelopoulos, Phys. Rev. Lett.
{\bf 62}, 2342 (1989).  
\item{4.} A. Gupta and T. Padmanabhan, Phys. Rev. {\bf D57},7241
(1998). An approach using non-locality has been proposed by
  B. Mashoon,Proc. VII Brazilian School of
Cosmology and Gravitation, Editions Fronti\'eres (1944);
  Phys. Lett. A {\bf 145}, 147 (1990); Phys. Rev.A{\bf 47}, 4498 (1993).
 We thank J. Beckenstein for bringing the latter works to our
  attention.
\item{5.} Y. Ashkenazy and L.P. Horwitz, chao-dyn/9905013, submitted.
\item{6.}   E.C.G. Stueckelberg, Helv. Phys. Acta {\bf 14}, 322
(1941); {\bf 14}, 588 (1941); JR.P. Feynman, Rev. Mod. Phys. {\bf 20},
 367 (1948); R.P. Feynman, Phys. Rev. {\bf 80}, 440(1950);
 J.S. Schwinger, Phys. Rev. {\bf 82}, 664 (1951);L.P. Horwitz and C. Piron,
 Helv. Phys. Acta {\bf 46}, 316 (1973).
\item{7.}  J.T. Mendon\c ca and L. Oliveira e Silva, Phys. Rev E {\bf
55}, 1217 (1997).
\item{8.} D. Saad, L.P. Horwitz and R.I. Arshansky, Found. of
Phys. {\bf 19}, 1125 (1989); M.C. Land, N. Shnerb and L.P. Horwitz,
Jour. Math. Phys. {\bf 36}, 3263 (1995); N. Shnerb and L.P. Horwitz,
Phys. Rev A{\bf 48}, 4068 (1993). We use a different convention for
the parameters here.
\item{9.} See, for example, J.D. Jackson, {\it Classical
Electrodynamics\/},  2nd
edition, John Wiley and Sons, New York(1975); F. Rohrlich, {\it
Classical Charged Particles}, Addison-Wesley,
 Reading, (1965); S. Weinberg, {\it Gravitation and Cosmology:
Principles and Applications of the General Theory of Relativity},
Wiley, N.Y. (1972).
\item{10.}M.C. Land and L.P. Horwitz, Found. Phys. Lett. {\bf 4}, 61
(1991); M.C. Land, N. Shnerb and L.P. Horwitz, Jour. Math. Phys. {\bf
        36}, 3263 (1995).
\item{11.} For example,   A.A. Sokolov and I.M. Ternov, {\it Radiation from
Relativistic Electrons}, Amer. Inst. of
Phys. Translation Series, New York (1986).
\item{12.}  P.A.M. Dirac, Proc. Roy. Soc. London Ser. A, {\bf 167},
148(1938).  
\item{13.}M.C. Land and L.P. Horwitz, Found. Phys. {\bf 21}, 299
(1991).
\item{14.} L.D. Landau and E.M. Lifshitz,{\it The Classical Theory of
Fields} 4th ed., (Pergamon Pr., Oxford, 1975).
\item{15.} A.O. Barut and Nuri Unal, Phys. Rev {\bf A40}, 5404 (1989) found 
non-vanishing contributions of the type ${\dot x}_\nu {\ddot x}^\nu$ to the
 Lorentz-Dirac equation in the presence of spin.
                              
\vfill
\end
\bye